\documentclass[%
aip,graphicx,amsmath,amssymb,preprint]{revtex4-1}
\usepackage{graphicx,epsfig,fancyhdr,epsf,txfonts,epstopdf}
\usepackage{lineno}
\begin{document}


\title{An alternative form of the fundamental plasma emission through the coalescence of Z-mode waves with whistlers} 


\author{Sulan Ni}
\affiliation{Institute of Space Sciences, Shandong University, Weihai, Shandong 264209, People’s Republic of China}
\affiliation{Center for Integrated Research on Space Science, Astronomy, and Physics, Institute of Frontier and Interdisciplinary Science, Shandong University, Qingdao, Shandong, 266237, People’s Republic of China}

\author{Yao Chen}
\email{yaochen@sdu.edu.cn}
\affiliation{Institute of Space Sciences, Shandong University, Weihai, Shandong 264209, People’s Republic of China}
\affiliation{Center for Integrated Research on Space Science, Astronomy, and Physics, Institute of Frontier and Interdisciplinary Science, Shandong University, Qingdao, Shandong, 266237, People’s Republic of China}

\author{Chuanyang Li}
\affiliation{Institute of Space Sciences, Shandong University, Weihai, Shandong 264209, People’s Republic of China}
\affiliation{Center for Integrated Research on Space Science, Astronomy, and Physics, Institute of Frontier and Interdisciplinary Science, Shandong University, Qingdao, Shandong, 266237, People’s Republic of China}

\author{Jicheng Sun}
\affiliation{Department of Physics, Auburn University, Auburn, AL, 36849, USA }

\author{Hao Ning}
\affiliation{Institute of Space Sciences, Shandong University, Weihai, Shandong 264209, People’s Republic of China}
\affiliation{Center for Integrated Research on Space Science, Astronomy, and Physics, Institute of Frontier and Interdisciplinary Science, Shandong University, Qingdao, Shandong, 266237, People’s Republic of China}

\author{Zilong Zhang}
\affiliation{Institute of Space Sciences, Shandong University, Weihai, Shandong 264209, People’s Republic of China}
\affiliation{Center for Integrated Research on Space Science, Astronomy, and Physics, Institute of Frontier and Interdisciplinary Science, Shandong University, Qingdao, Shandong, 266237, People’s Republic of China}


\date{\today}

\begin{abstract}
Plasma emission (PE), i.e., electromagnetic radiation at the plasma frequency and its second harmonic, is a general process occurring in both astrophysical and laboratory plasmas. The prevailing theory presents a multi-stage process attributed to the resonant coupling of beam-excited Langmuir waves with ion-acoustic waves. Here we examine another possibility of the fundamental PE induced by the resonant coupling of Z-mode and whistler (W) waves. Earlier studies have been controversial in the plausibility and significance of such process in plasmas. In this study we show that the matching condition of three wave resonant interaction (Z+W $\rightarrow$ O) can be satisfied over a wide regime of parameters based on the magnetoionic theory, demonstrate the occurrence of such process and further evaluate the rate of energy conversion from the pumped Z or W mode to the fundamental O mode with particle-in-cell (PIC) simulations of wave pumping. The study presents an alternative form of the fundamental PE, which could possibly play a role in various astrophysical and laboratory scenarios with both Z and W modes readily excited through the electron cyclotron maser instability.
\end{abstract}


\maketitle 

Plasma emission (PE) is loosely defined as electromagnetic radiations at the plasma frequency and its second harmonic, corresponding to the fundamental (F) and harmonic (H) emissions. The process accounts for coherent radiations observed from a wide range of astrophysical and laboratory plasmas, including but not limited to (1) the solar atmosphere and the interplanetary space, such as type-III and II radio bursts \cite{1985_McLean,1984_Dulk}, (2) the planetary electron foreshocks such as those of the Jupiter and the Earth \cite{1984_Etcheto,1984_Moses,1987_Cairns}, (3) the outer heliosphere \cite{1984_Kurth}, (4) flare stars \cite{1985_Kuijpers,2014_Loeb}, and (5) laboratory experiments \cite{1984_Intrator,1985_Whelan}.

The prevailing theory for PE presents a multi-stage nonlinear process, starting from the kinetic bump-in-tail instability driven by beams of energetic electrons, scattering of the excited Langmuir waves by ion-acoustic waves to account for the F radiation, and coalescence of forward- and backward-propagating Langmuir waves to yield the H radiation. The framework of PE was originally suggested by Ginzburg and Zheleznyakov \cite{1958_Ginzburg}, being developed and improved substantially over the decades \cite{1980_Melrose,1987_Cairns,1994_Robinson}. Latest efforts are mainly along three fronts: (1) developing the electromagnetic weak-turbulence theory to describe the nonlinear evolution of the beam-plasma instability \cite{Ziebell_2014,Ziebell_2015,Lee_2019}, (2) verifying the complete multi-step nonlinear PE process driven by single- or counter-propagating beam using the full electromagnetic PIC simulation \cite{Ganse_2012,Thurgood_2015,Che_2017,Henri_2019}, (3) simulating major characteristics of various types of radio bursts observed in space, such as the solar type-II and -III bursts, by incorporating the reduced PE theory into large-scale models \cite{Schmidt_2012a,Schmidt_2012b,Ratcliffe_2014a,Ratcliffe_2014b,Reid_2014}.


For other possible forms of PE, in a recent study Ni et al. \cite{Ni_2020} suggested that the F emission could be a result of the resonant coalescence of magnetoionic Z and whistler (W) modes, and the H emission be a result of the coalescence of upper hybrid (UH) waves, in weakly magnetized plasmas. These modes are excited by energetic electrons of the Dory-Guest-Harris (DGH \cite{1965_Dory}) distribution through electron cyclotron maser instability (ECMI), in a parameter regime with a large ratio of plasma frequency to electron gyro-frequency ($\omega_{pe}/ \Omega_{ce}$ (=10)). The DGH distribution is of the double-sided loss cone type, usually applied for energetic electrons trapped within a magnetic structure. The essential nature of ECMI is identical to the alternative type of coherent emission in plasmas with a small value of $\omega_{pe}/ \Omega_{ce}$ ($<<1$), known as the electron cyclotron maser emission (ECME \cite{1958_Twiss,1979_Wu}).

Plasma radiation involving whistlers was first suggested by Chiu \cite{1970_Chiu} and Chin \cite{1972_Chin} through their coalescence with Langmuir waves. Melrose \cite{1975_Melrose} rejected the idea since (1)  for thermal Langmuir waves, as originally proposed by Chiu, the coupling cannot explain the high brightness of coherent radiation, (2) for beam-excited Langmuir waves the wave number ($k$) is approximately $\omega_{pe}/v_b$ ($>>\omega_{pe}/c$) too large to meet the conditions of coupling with whistlers whose $k$ is usually less than $\omega_{pe}/c$, while the F emission of O mode usually has a small $k$ ($<<\omega_{pe}/c$), where $v_b$ ($<<c$) represents the beam velocity. He also concluded that the only wave around the plasma frequency that could coalesce with whistlers is the superluminal Z mode that is electromagnetic with a phase speed close to or larger than $c$ and a short-enough wave number.

Dispersion relations of Z mode and whistlers can be approximated by the magnetoionic theory \cite{1931_Hartree,1932_Appleton}. Using simplified expressions, Melrose \cite{1975_Melrose} deduced that the Z-W resonant coupling can occur only under very restrictive condition and therefore PE involving whistlers is unlikely to be important, at least not in coronal plasmas. Despite this negative statement, some later studies still assumed the involvement of whistlers to generate the fundamental emission. For example, Takakura \cite{1981_Takakura} and Mann et al. \cite{1987_Mann} suggested the coalescence of whistlers and plasma waves which are scattered by thermal ions to reach a small wave number, to account for the F emission of Type-III solar radio bursts. Whistlers are also used to explain formations of fine spectral structures of Type-IV solar radio bursts through rapid diffusion of energetic electrons (e.g., \cite{1976_Chernov,2006_Chernov}). Recently, Ni et al. \cite{Ni_2020} reported the apparent occurrence of the three-wave interaction (Z+W$\rightarrow$O) according to PIC simulations. Yet, plausibility and significance of the Z-W-O interaction in plasma radiation process have not been carefully examined, and direct solution of the matching conditions and proof of the occurrence of coalescence are not available.

In this Letter, we first present solutions of the matching conditions of three-wave resonant coupling (Z+W$\rightarrow$O) using the dispersion relations given by the canonical magnetoionic theory, which describes the propagation of plasma waves in cold magnetized electron plasmas. The relations can be found in many textbooks of plasmas and will not be repeated here. The matching conditions can be written as
\begin{equation}
\omega_Z + \omega_W = \omega_O, \ \ \ \ \vec {k}_Z +\vec {k}_W = \vec {k}_O
\end{equation}
where subscripts represent corresponding wave modes. Then we demonstrate the occurrence of the three-wave coupling process by pumping waves of Z and W modes, either separately or together, into the PIC system that is simulated with the open-source Vector PIC (VPIC) code provided by the Los Alamos National Labs and run on the supercomputers of Beijing Super Cloud Computing Center (BSCC). It is a fully-kinetic electromagnetic and relativistic PIC code \cite{2008a_Bowers,2008b_Bowers,2009_Bowers}, run in two spatial dimensions ($x, z$) with three velocity components using periodic boundary conditions.
Note that the main purpose of the present study is to verify whether the matching conditions can be met by solving them directly and confirm the occurrence of the matching process using PIC simulations, rather than simulating the complete ECMI-plasma emission process, which has been done in Ni et al. \cite{Ni_2020} and Li et al. \cite{Li_2021}

To solve the resonance conditions for the specific case with $\omega_{pe}/\Omega_{ce} = 10$, we vary the propagation angle ($\theta_{kB}$) for Z and W modes from  0$^\circ$ to 360$^\circ$. For each angle we obtain the corresponding cutoff- and resonance- frequencies which are then used to limit the ranges of possible frequency ($\omega$) of the two modes. We put 500 grid points in each range. By solving the magnetoionic dispersion relations, we obtain values of $k$ corresponding to each combination of $\omega$ and $\theta_{kB}$ for Z and W modes, ending up with $500^4$ sets of [$\omega$, $k$] combinations of wave solutions. The units of $\omega$ and $k$ are taken to be $\Omega_{ce}$ and $\Omega_{ce}/c$, respectively. The parameters are then summed up according to Equation (1), if the sums meet the magnetoionic O-mode dispersion relation with an accuracy of $<10^{-4}$, then we save the parameter set as a resonance solution. In total, we obtain 14285 such solutions. We have tried many other cases with $\omega_{pe}/\Omega_{ce}$ ranging from 3 to 15, getting similar results.

We show 7 examples in Figure 1, with wave vectors shown in panel (a) and frequencies plotted onto the corresponding magnetoionic dispersion curves in panels (b) and (c). These examples distribute within the range of [$0^\circ, 90^\circ$] for the Z-mode propagation angle. Since the whistlers experience strong damping when reaching sufficiently-high frequency ($\sim >\Omega_{ce}/2$), we do not consider the possibility of Z and W coalescence into the X mode which has a cutoff frequency around 10.5 $\Omega_{ce}$. We see that the coalescing Z and W modes distribute within a relatively large range of $\omega$, $k$, and $\theta_{kB}$. This can be confirmed from Figure 2, in which we plot the histogram distribution of solution numbers for the three modes. The $k$ values of both Z and W modes distribute mainly within a range of [2.5, 9] $\Omega_{ce}/c$, the O mode distributes mainly within a range of [0, 4] $\Omega_{ce}/c$; the frequencies of Z and W lie mainly in ranges of [9.8, 10.0] $\Omega_{ce}$ and [0, 0.4] $\Omega_{ce}$ respectively, and their propagation angles present a preference of distribution over the parallel to oblique propagation; for the O mode the frequency range is mainly in the range of [10, 10.4] $\Omega_{ce}$ while its propagation angle manifests a quasi-isotropic distribution with a weaker preference over the parallel and anti-parallel propagation. These solutions show that the three-wave coupling process (Z+W$\rightarrow$O) is quite general and takes place over a broad range of wave parameters, as seen from the perspective of resonance conditions.

To demonstrate the actual occurrence of the Z+W$\rightarrow$O coalescence, we develop a wave-pumping technique. Different wave properties, such as components of the electro-magnetic fields and the velocity, are related linearly to each other. The relationships can be easily deduced from the mass and momentum equations for cold-plasma electrons and the Maxwell equations. All wave properties vary in space in form of cosine function with a prescribed initial phase angle ($\sim \cos (\vec k \cdot \vec r + \psi_0)$, where $\vec r = x \hat e_{x} + z \hat e_{z}$). Revising the initial conditions of the PIC simulation accordingly allows us to pump various wave modes into the system.

In the PIC simulation, the plasma consists of equal number of thermal electrons and protons with Maxwellian distribution of $2$ MK, typical temperatures for solar coronal plasmas; $\omega_{pe} / \Omega_{ce}$ is taken to be $10$, the background magnetic field $\textbf{B}_0 = B_0 \hat{e}_z$ is uniform and along the $z$ direction, the wave vector $\vec k$ is in the $xOz$ plane. We employ 1000 macro-particles per species per cell. The simulation lasts for 2000 $\omega_{pe}^{-1}$ within the spatial domain of $2048 \Delta \times 2048 \Delta$, where $\Delta=3.25\lambda_D$ and $\lambda_D$ is the Debye length of thermal electrons. The time step is set in order to ensure the Courant conditions $\Delta t = 0.029$ $\omega_{pe}^{-1}$. A realistic proton-to-electron mass ratio of 1836 is used.

The resonance solution with $\theta_{kB} = 45^\circ$ of the Z mode (see the red arrows plotted in Figure 1 for this case and exact values of relevant parameters in its caption) is selected for further PIC simulation. We pump Z and W modes with corresponding parameters into the PIC system, either separately or together, obtaining three cases (referred to as Cases Z, W, and Z+W). Comparing them, we can tell whether the three-wave resonance indeed takes place and evaluate the percentage of energy that can be converted from Z or W to O.

The distribution of the maximum spectral intensity in the $\vec k$ space for the three cases is illustrated in Figure 3, and the $(\omega, k)$ dispersion curves along $\theta_{kB} = 45^\circ$ and $224^\circ$ are presented in Figure 4 (and the accompanying movie for other relevant propagation angles) (Multimedia view) superposed by those of magnetoionic modes. Various modes (Z, W, and O) can be easily identified. The three sets of arrows in Figure 3 point to parameters given by the above-selected resonance solution. They also point right at the center of areas with enhanced intensity of the PIC simulation. This indicates that the modes have been pumped correctly. The Z mode is dominated by the electric field fluctuation while the W mode is dominated by the magnetic field fluctuation, as expected. The obtained O mode is dominated by the parallel component of electric field ($E_z$) since its frequency is at or only slightly larger than $\omega_{pe}$. Note that the simulated wave characteristics distribute over a relatively broad angular region of $\sim 20^\circ$ (see the accompanying movie), therefore the characteristics of both Z and O modes at $45^\circ$ are almost the same as those at $47.7^\circ$, so in Figure 4 (Multimedia view) only diagrams at $45^\circ$ are shown for these two modes.

It should be highlighted that the O mode appears with the expected wave vector only in Case Z+W, confirming the occurrence of the Z+W$\rightarrow$O coalescence, and the maximum of the O-mode intensity is centered around the parameters prescribed by the selected coalescence solution. This presents unambiguous evidence of the occurrence of the three-wave interaction of Z+W$\rightarrow$O.

The temporal profiles of energy of the three modes are plotted in Figure 5a for the three cases, calculated using the Parseval's theorem. For Z and W modes, the total electro-magnetic field energy is calculated directly from the PIC simulation, while for the O mode the total energy is calculated by the simulated magnetic energy multiplied by the theoretical magnetoionic ratio of the total to magnetic energy, to avoid interference of nearby Z mode.

For Case Z (W), the energy profiles of W (Z) and O modes represent the corresponding noise levels. The input energy of Z (W) modes are above the noise levels by about 2-3 orders in magnitude, and the obtained O mode energy in Case Z+W is above the corresponding noise level by one order in magnitude. The input energy of Z and W modes is taken to be close to each other, referred to as $E_{Z,W}^0 \sim 5.8\times 10^{-6} E_{B_0}$ hereinafter, where $E_{B_0}$ represents the total energy of the background magnetic field. Note that the pumped Z and W modes in all cases presented here should be regarded as small-amplitude waves since their energy are negligible in comparison to that of the background magnetic field. With time going on, energies of the pumped modes present an overall gradual decline, while the O mode energy increases gradually with time during the early stage of the interaction. After $t=750$ $\omega^{-1}_{pe}$, the energy reaches an asymptotic value of $\sim 0.5\% E_{Z,W}^0$, which gives the energy conversion rate of the PE process, from the pumped wave modes to the escaping radiation mode, in the present Z+W case. At this time ($t=750$ $\omega_{pe}^{-1}$), the relative decline of the Z mode energy in both Case Z and Case Z+W is about $\sim$10\%, while that of the W mode energy in both Case W and Case Z+W is about $\sim$20\%. Thus, the decline of energy due to the conversion to O mode in Case Z+W is minor, indicating that the decline of Z and W mode energies are mainly due to the damping of individual mode, rather than their interaction.

To investigate the variation of the conversion rate versus the pumped energy, we gradually increase the pumped energy of one mode of Z and W while fixing that of the other mode, taking Case Z+W to be the reference solution. This yields two sets of parameter studies, referred to as Set Z with Z-mode energy increasing and Set W with W-mode energy increasing. The results and corresponding lines of linear and polytropic fittings are plotted in Figure 5b. If the pumped energy is less than $8\times10^{-4}$ $E_{B0}$ or the obtained O-mode energy is less than ($0.2-0.3$ $E^0_{Z,W}$), then the O-mode energy ($E_O$) increases almost linearly with the input energy of either Z or W mode, while the O mode energy approaches an asymptotic value ($E_c$) of about 2$\times 10^{-6} E_{B0}$ ($\sim 0.3$ $E^0_{Z,W}$) if the pumped energy gets larger. The linear fitting lines for the two sets (Z and W) are respectively expressed as
$$
y=0.0025 x + 5.37\times10^{-9}, \ \ \ \ y=0.0021 x + 3.63\times10^{-8}
$$
The slope of the fitting lines can be regarded as the energy conversion rate from the pumped Z (for Set Z) or W (for Set W) mode to O mode for the respective linear-correlation regime.

These results, in particular the linear correlation found above, provide additional yet strong support to the occurrence of three-wave interaction with the O mode being a result of the nonlinear coalescence of Z and W. If further increasing the input energy above the levels of Sets Z and W, there appears unexpected interfering signals caused by numerical artifacts and/or nonlinear effect which make the uncertainty of mode evaluations larger. This is partially due to the usage of cold-plasma magnetoionic approximation in our wave-pumping method. Note that since the amount of energy pumped into the system is rather small, no obvious heating or acceleration of electrons are observed, and the original Maxwellian distribution remains.

To summarize, the Letter explored an alternative form of the fundamental plasma emission involving the nonlinear resonant coupling of Z-mode waves and whistlers. On the basis of cold-plasma magnetoionic theory of plasma waves, we solved the matching conditions of three-wave interaction of Z mode, whistlers, and O mode. It was found that the conditions can be satisfied within a quite general and broad regime of wave parameters. In addition, we developed a wave-pumping technique using the VPIC code to simulate the nonlinear coupling of Z and W modes and generation of the O-mode emission. By case and parameter studies, we presented solid evidence of the occurrence of the process and investigated the energy conversion rate from the pumped Z-mode waves and whistlers to the fundamental O-mode emission. It was found that there exists a critical value which determines how the coalesced O-mode energy varies according to the input energy of Z and W modes, either linearly below a certain threshold or asymptotically beyond. The maximum energy that can be converted to O mode was found to be about 30\% of the Z- or W- mode energy, whichever is smaller.

The prevailing theory of the fundamental PE is the scattering of beam-excited Langmuir waves by ion-acoustic wave or density inhomogeneities. The process discussed here is an alternative form of the fundamental PE. For this form to play a role in plasmas, both Z-mode waves and whistlers should be excited efficiently. This can be done through the ECMI driven by energetic electrons with velocity distribution functions of positive gradient along the perpendicular direction, i.e., with ${\partial f \over \partial v_{\perp}}>0$. Thus, the complete emission process can be referred to as the ECMI-induced PE. Examples of such distributions include the well-known loss cone distribution, and the DGH, ring, ring-beam, and horseshoe distributions. The first example of the ECMI-induced PE driven by DGH electrons has been presented by Ni et al. \cite{Ni_2020}.

Here we briefly list a few types of solar radio bursts as examples, to which the mechanism may be important, (1) type-IV bursts which are believed to be generated by energetic electrons trapped within eruptive magnetic structures (e.g., \cite{2002_Benz,2016_Vasanth,2019_Vasanth}), therefore loss-cone like distribution shall develop and play a role by exciting both W and Z-mode waves which then coalesce to yield the F emission; (2) type-I bursts which are likely related to electrons released into coronal loops by small-scale reconnections (see \cite{2017_Li}) and the form of electron distribution may be different from a beam type; (3) type-II bursts which have been assumed to be similar to type-IIIs in emission mechanism, yet doubts remain regarding the exact form of electron distributions, it is likely that type-IIs are excited by electrons that are trapped and scattered around the shock, thus for the same reason as above the mechanism discussed here may be significant \cite{2013_Chen,2014_Chen,2015_Kong}. Further investigations are demanded to clarify the role of the ECMI-induced PE mechanism in various circumstances of astrophysical and laboratory plasmas.

 This study is supported by the National Natural Science Foundation of China (11790303 (11790300), 11750110424, and 11873036). The authors acknowledge the Beijing Super Cloud Computing Center (BSCC, URL: http://www.blsc.cn/) for providing high-performance computing (HPC) resources, and the open-source Vector-PIC (VPIC) code provided by Los Alamos National Labs (LANL). The authors acknowledge the helpful discussions with  Drs. Xin Tao, Quanming Lu, and  Xinliang Gao at USTC, and Dr. Xiaocan Li at LANL.

\section{Data Availability} 
The data that support the findings of this study are available from the corresponding author upon reasonable request.

\begin{figure}
	\includegraphics[width=0.99\linewidth]{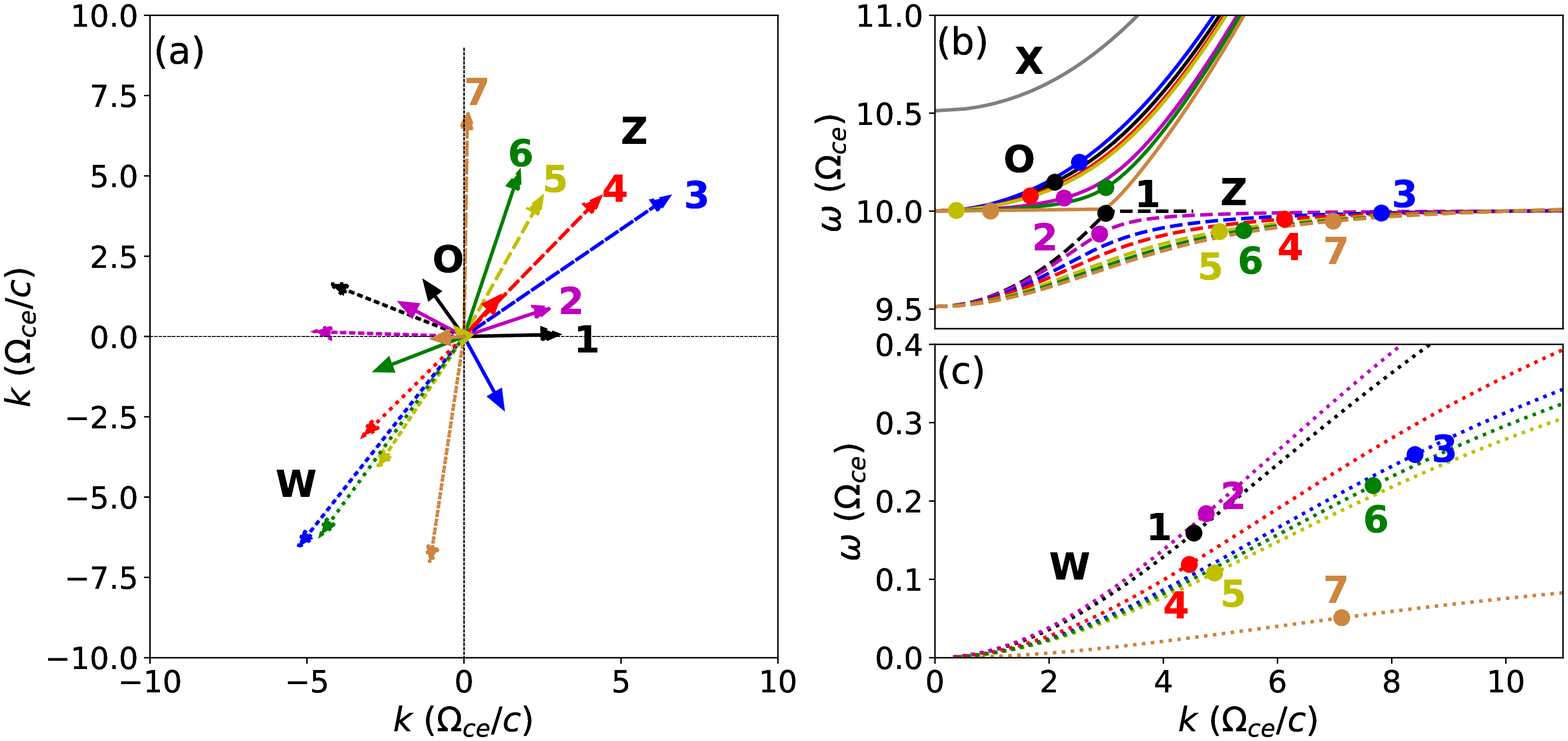}%
	\caption{Seven examples selected from the solutions of resonance conditions.
		(a) Wave vectors ($\vec k$) of Z, W, and O, (b-c) frequencies ($\omega$) of Z, O, and W plotted onto the corresponding magnetoionic dispersion curves (dashed lines for Z mode, dotted lines for W mode, and solid lines for O mode), the dispersion curve of X mode along $\theta_{kB} = 45^{\circ}$ is also plotted in panel (b). The selected wave mode for further PIC simulation is labeled by `4', exact values of ($\omega$, $\vec k$, $\theta_{kB}$) for this solution are (9.959 $\Omega_{ce}$, 6.123 $\Omega_{ce}/c$, 45$^{\circ}$) for Z mode, (0.119 $\Omega_{ce}$, 4.455 $\Omega_{ce}/c$, 224$^{\circ}$) for W mode, and (10.078 $\Omega_{ce}$, 1.671 $\Omega_{ce}/c$, 47.7$^{\circ}$) for O mode.}
	\label{1}
\end{figure}

\begin{figure}
	\includegraphics[width=0.99\linewidth]{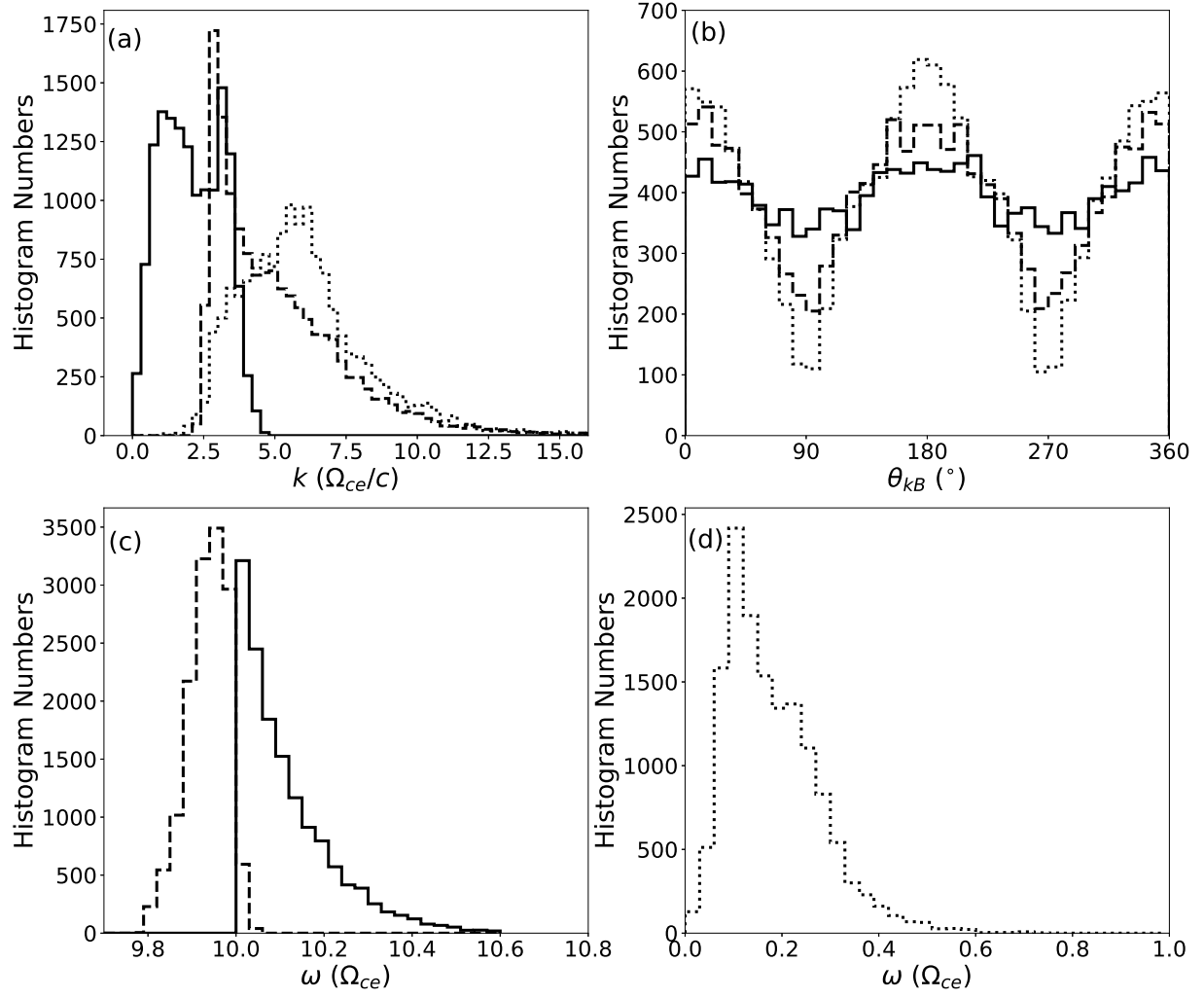}%
	\caption{Histogram distributions of solutions of the matching condition (see Equation (1)) of three wave resonant interaction (Z+W $\rightarrow$ O): (a) for wave numbers , (b) for propagating angles, and (c-d) for frequencies.}
	\label{2}
\end{figure}
\begin{figure}
	\includegraphics[width=0.99\linewidth]{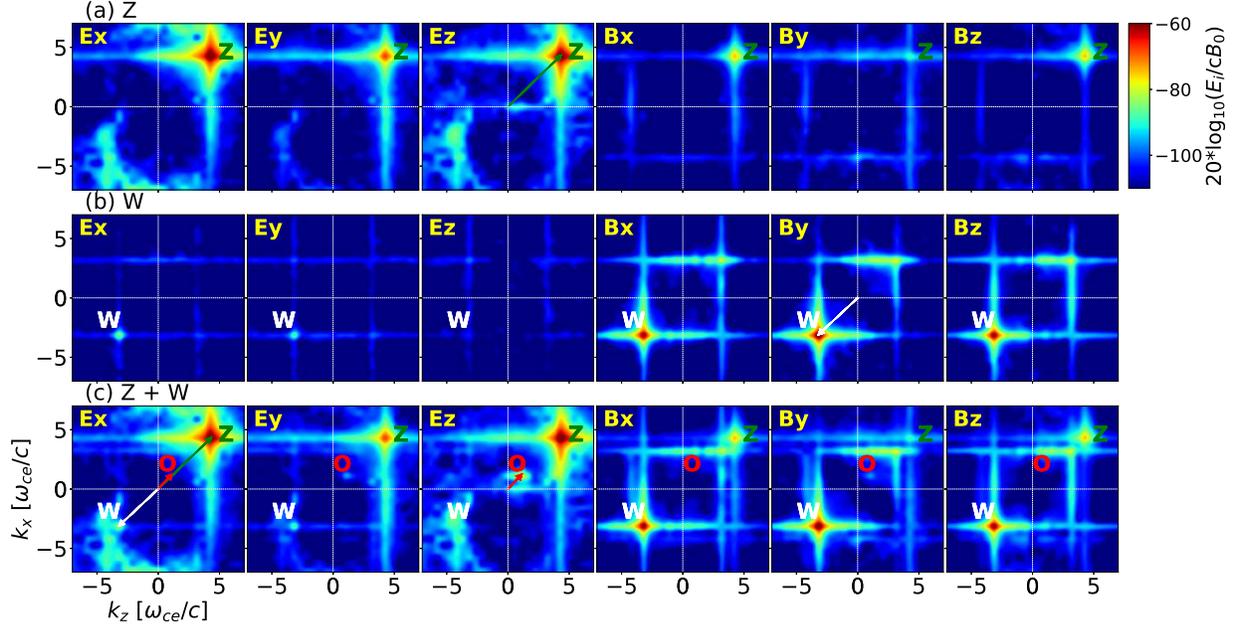}%
	\caption{Maximum intensity distributions of the six field components in the $\vec k$ space for Case Z (a), Case W (b), and Case Z+W (c), as shown by the colormap of 20 $log_{10}$ [($E_x$, $E_y$, $E_z$, $B_x$, $B_y$, $B_z-B_0$)/($cB_0$)]. All panels use the same colorbar as shown next to panel (a). Wave vectors of the three modes corresponding to the selected solution (see solution `4' plotted in Figure 1(a)) are plotted as arrows.}
	\label{3}
\end{figure}
\begin{figure}
	\includegraphics[width=0.99\linewidth]{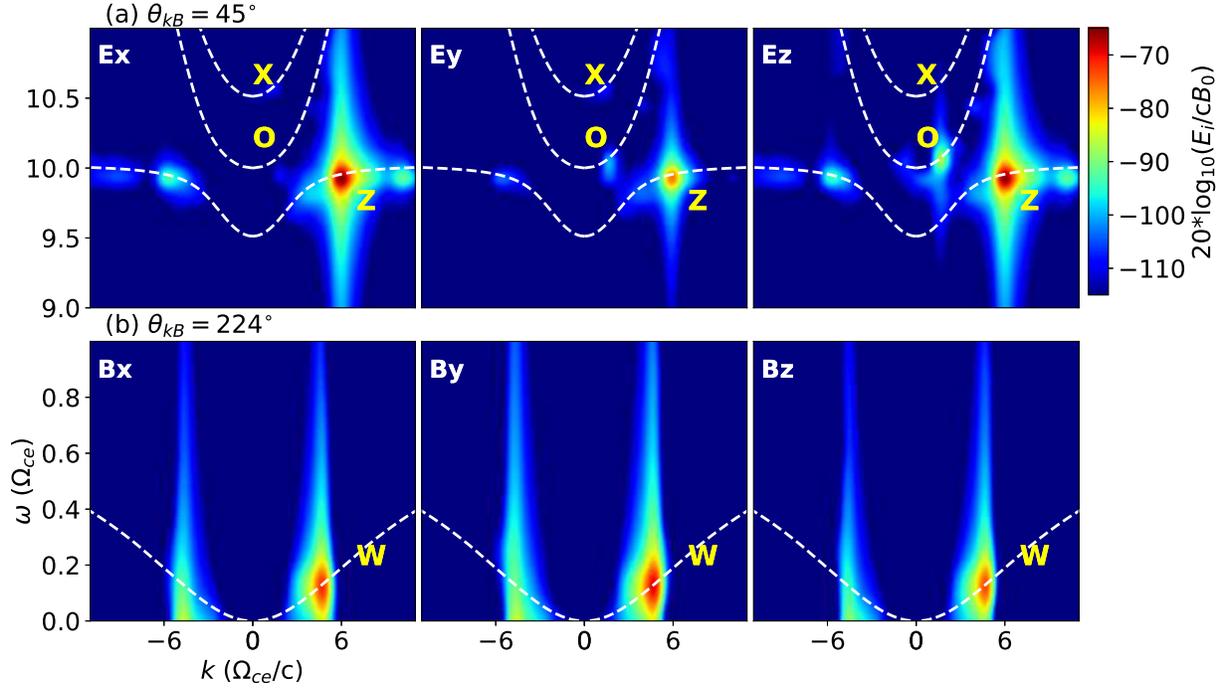}%
	\caption{Wave dispersion diagrams of (a) $E_x$, $E_y$, $E_z$ along $\theta_{kB} = 45^{\circ}$ and (b) $B_x$, $B_y$, $B_z$ along $\theta_{kB} = 224^{\circ}$. The corresponding dispersion curves of Z, W, O, and X modes are also plotted. An accompanying animation is available online (Multimedia view).}
	\label{4}
\end{figure}
\begin{figure}
	\includegraphics[width=0.99\linewidth]{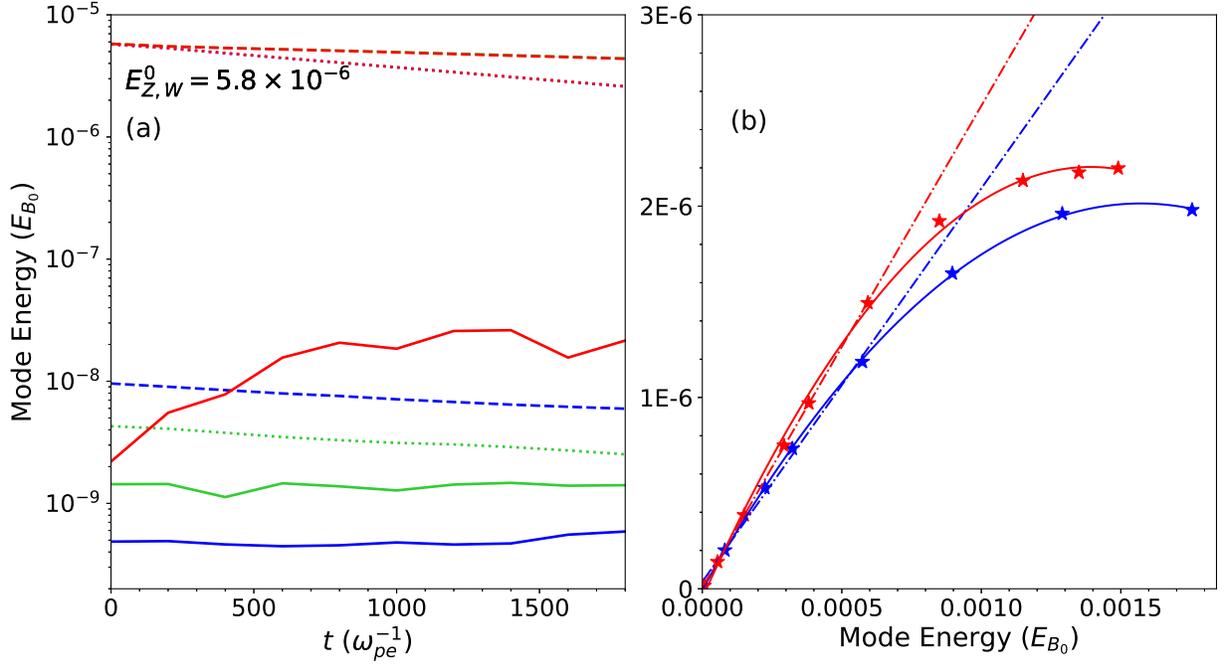}%
	\caption{(a) Temporal profiles of energy of Z mode (dashed), W mode (dotted), and O mode (solid) for Case Z (green), Case W (blue), and Case Z+W (red). The energy is normalized with the total energy of the background magnetic field ($E_{B_0}$), (b) variation of O-mode energy with the increase of the pumped energy of Z and W modes (red for Set Z and blue for Set W) with stars representing results of the two sets of parameter study. Dot-dashed and solid lines represent linear and polytropic fitting curves of the PIC results, respectively.}
	\label{5}
\end{figure}

\end{document}